# Community recommendations on cryoEM data archiving and validation

*Outcomes of a wwPDB/EMDB workshop on cryoEM data management, deposition and validation*

## The Authors

(Names and affiliations at the end of this document)

## Abstract

In January 2020, a workshop was held at EMBL-EBI (Hinxton, UK) to discuss data requirements for deposition and validation of cryoEM structures, with a focus on single-particle analysis. The meeting was attended by 47 experts in data processing, model building and refinement, validation, and archiving of such structures. This report describes the workshop's motivation and history, the topics discussed, and consensus recommendations resulting from the workshop. Some challenges for future methods-development efforts in this area are also highlighted, as is the implementation to date of some of the recommendations.

## Synopsis

This white paper describes recommendations to wwPDB, EMDB and the cryoEM community regarding archiving and validation of single-particle structures and volumes.

## Abbreviations

2D, two-dimensional
3D, three-dimensional
3DEM, 3D Electron Microscopy
BMRB, Biological Magnetic Resonance Data Bank
cryoEM, cryogenic-specimen Electron Microscopy
EM, Electron Microscopy
EMBL, European Molecular Biology Laboratory
EMBL-EBI, EMBL-European Bioinformatics Institute
EMDB, Electron Microscopy Data Bank
EMPIAR, Electron Microscopy Public Image Archive
EMDR, Electron Microscopy Data Resource
FSC, Fourier-Shell Correlation
IHM, Integrative/Hybrid Methods
MX, Macromolecular Crystallography
NCMI, National Center for Macromolecular Imaging
NIGMS, National Institute of General Medical Sciences
NIH, National Institutes of Health
NMR, Nuclear Magnetic Resonance spectroscopy
OneDep, wwPDB unified system for deposition of atomic structures and experimental data and metadata for biological macromolecules
PDB, Protein Data Bank
PDBe, Protein Data Bank in Europe
PDBj, Protein Data Bank Japan
RAPS, Rotationally-Averaged Power Spectrum
RCSB, Research Collaboratory for Structural Bioinformatics
RCSB PDB, RCSB Protein Data Bank
RSRZ, Real-Space R-value Z-score
SASBDB, Small-Angle Scattering Biological Data Bank
SPA, Single-particle analysis
STA, Subtomogram averaging
VTF, Validation Task Force
wwPDB, Worldwide Protein Data Bank

# Introduction & background

Structural biology, the study of the 3D structures of biological entities on scales from biomolecules to cells, has had an enormous impact on our understanding of biology and biological processes in health and disease. For many years, single-crystal X-ray diffraction was the main technique used to obtain 3D structures of biological macromolecules with (near-)atomic detail. Since the 1980s, NMR spectroscopy techniques have also contributed thousands of structures, albeit largely limited to relatively small (and soluble) molecules. Electron diffraction was already used in the 1970s to investigate the structure of an integral membrane protein, bacteriorhodopsin, although an atomic model was not described until 1990 (Henderson *et al.*, 1990). During the 1980s, pioneers in the cryoEM field developed experimental and computational methods (notably, embedding specimens in vitreous ice, low-dose electron imaging and detection, and particle reconstruction from projection images) which enabled increasingly high-resolution studies of a variety of biological specimens, and eventually, in what has been termed the "*resolution revolution*" (Kühlbrandt, 2014), atomistic modelling.

**Archiving.** The results of these structural studies by MX, NMR and 3DEM have been captured in the single global archive of atomic models of biomacromolecules and their complexes, the PDB (wwPDB consortium, 2019). With great foresight, the protein crystallography community established this archive in 1971 and this has turned out to have been a landmark event in the history of public archiving of open scientific data. The PDB was originally hosted at Brookhaven National Laboratory. Since the early 2000s, the PDB has been managed and operated through the wwPDB, a collaboration involving five partners across the USA, Europe, and Japan (Berman *et al.*, 2003). In addition to atomic models, the PDB also captures some derived experimental data, namely crystallographic structure factors and NMR chemical shifts and restraints. In 2002, another community initiative led to the establishment of the Electron Microscopy Data Bank (EMDB) (Tagari *et al.*, 2002). EMDB archives processed experimental data from a variety of 3DEM modalities, most notably single-particle analysis (SPA), electron tomography and subtomogram averaging (STA), helical reconstruction (HR), and electron crystallography (EC) (wwPDB consortium, 2024). These modalities all produce data from which ultimately 3D volumes can be determined. Deposition of atomic 3DEM structural models to the PDB occurs through the wwPDB OneDep System (Young *et al.*, 2017), which also supports deposition of 3DEM processed experimental data (predominantly volumes) and metadata to EMDB. Between 2007 and 2020, EMDB was operated jointly by EMDR, a collaborative project between EMBL-EBI in the UK and RCSB and NCMI in the USA. As of 2021, EMDB is a core wwPDB archive and a full wwPDB member. Hence, the wwPDB partners now jointly manage deposition, validation, and biocuration of all 3DEM atomic structural models, processed experimental data, and metadata.

**Need for validation.** In the late 1980s, some protein crystallographers began to realise that the structural models they produced were of varying reliability (and in rare cases completely wrong), particularly when the resolution of the experimental data was low (Brändén & Jones, 1990). This realisation led to the development of (i) new validation techniques, such as statistical cross-validation, and new metrics such as the free R-value (Brünger, 1992) and real-space R-value (Jones *et al.*, 1991), (ii) new validation software, either embedded in model-building software such as O (Jones *et al.*, 1991), or as stand-alone packages such as ProCheck (Laskowski *et al.*, 1993), WHATCHECK (Hooft *et al.*, 1996), and MolProbity (Davis *et al.*, 2004), and (iii) recommendations about "good practice" to minimise the likelihood of serious errors going undetected and making it into final models, the PDB and the literature (Kleywegt & Jones, 1995) (Kleywegt & Jones, 1997).

In 2006, three models of human C3b complement-system components from different laboratories were published back-to-back in *Nature*, and structure comparison revealed one model to be an outlier with some physically unlikely features (Abdul Ajees *et al.*, 2006). This led to suspicions of scientific fraud and data fabrication, not just for this model but for about a dozen structures published by the PI responsible for the deviating model (Borrell, 2009), and these suspicions were later confirmed following a thorough investigation by the US Office for Research Integrity (https://ori.hhs.gov/content/case-summary-murthy-krishna-hm). This case sparked widespread concern in the structural biology community and directly led to the deposition of crystallographic structure factors (2008) and NMR chemical shifts and restraints (2010) being made mandatory. The wwPDB leadership at the time realised that the new deposition requirements opened up entirely new opportunities to validate all models against the supporting experimental data at the time of deposition. A meeting was organised in 2008 of a newly established wwPDB X-ray Validation Task Force (VTF) which produced an influential report a few years later (Read *et al.*, 2011). It made wide-ranging recommendations concerning validation of MX structures, including detailed suggestions on how the validation results should be reported. These recommendations led to the development of the now familiar wwPDB validation reports, which have been generated for X-ray structure depositions since 2013 and later were also provided for all legacy entries in the archive (Gore *et al.*, 2012) (Gore *et al.*, 2017). A similar meeting of NMR experts in 2009 led to a report and recommendations for validating NMR models and data as well (Montelione *et al.*, 2013), and validation reports for NMR entries have been available since 2016.

**Validation and 3DEM.** In 2010, as part of the NIH-NIGMS-funded EMDR project mentioned earlier, an EM VTF meeting was organised and its recommendations were published in 2012 (Henderson *et al.*, 2012). At that time, there were only ~1000 entries in EMDB (and only ~250 3DEM structural models in the PDB), and their resolution was generally relatively low and rarely allowed for an all-atom model to be constructed *de novo* and refined. The EM VTF report made some preliminary recommendations to the archives and also identified many areas in which further research and methods development by the 3DEM community were needed. A key recommendation was to establish two fully independent half-datasets at the outset for evaluating resolution by FSC of the resulting independent half-maps.

Since that meeting in 2010, there had been many developments in the field, which made it necessary to reconvene a group of experts to provide updated and more specific recommendations regarding the deposition and validation of 3DEM structures (maps and models). These developments include:
- The "*resolution revolution*" in cryoEM, which was enabled by direct electron detectors, improved microscopes (e.g., improved optical and mechanical stability, coherent electron source, and many improvements in collection efficiency), and better software to reconstruct the particles from the projection images. The result was a flood of 3DEM maps at resolutions sufficient for atomic structural models to be constructed and deposited to the PDB (**Figure 1**).
- The EMPIAR archive for raw 2D image data underpinning 3DEM volumes deposited to EMDB was established in 2013 (Iudin *et al.*, 2016) (Iudin *et al.*, 2023). This had been one of the recommendations of the 2010 EM VTF meeting and of later EMDB workshops (Patwardhan *et al.*, 2012) (Patwardhan *et al.*, 2014) and has enabled validation of published maps, as well as development and testing of new and improved methods for processing, analysing and validating cryoEM data.
- A number of challenge activities have been organised by the 3DEM community over the past two decades to compare existing methods and encourage further software development, including stimulating the development of new validation metrics (reviewed in (Lawson *et al.*, 2020)). The EMDR project has sponsored challenges to explore methods for generation and validation of coordinate models from 3DEM maps (Ludtke *et al.*, 2012) (Lawson & Chiu, 2018) (Lawson *et al.*, 2021), as well as methods for map reconstruction and validation (Lawson & Chiu, 2018). Additional community-organised challenge topics have included particle picking (Zhu *et al.*, 2004) and contrast-transfer-function correction (Marabini *et al.*, 2015).
- In January 2019, the UK EM Validation Network organised an expert workshop at EMBL-EBI on "*Frontiers in cryoEM Validation*". Where the EM VTF is expected to survey the field and make recommendations based on well-established science and broad community consensus, the *Frontiers* workshop identified needs of and challenges to the field, although many of its discussions fed into the 2020 workshop, and many of the participants attended both meetings.

**Validation reports.** 3DEM structural data is deposited to PDB and EMDB through a single system called OneDep (Young *et al.*, 2017), developed and maintained jointly by the wwPDB partners. Towards the end of the deposition process, a report is generated that summarises experimental metadata and provides validation of model and data (Gore *et al.*, 2017). Atomic model validation of 3DEM (and NMR) models follows the recommendations of the X-ray VTF.

A number of 3DEM-specific features were added to the reports shortly before the workshop, mainly based on content originally developed for the "visual analysis" web pages provided on the EMDB website (Lagerstedt *et al.*, 2013). In the case of a single-particle study with an atomic model, this included at the time:
- "Table 1", an overview of experimental details.
- Map visualisations, including orthogonal projections (along X, Y and Z), central slices, slices with the highest variance, and orthogonal surface views of the map (and also of any masks that were deposited) rendered at the contour level recommended by the depositor.
- Map-analysis graphs, including a histogram of map values (which reveals if parts of the map were masked), volume-estimate curve (enclosed volume as a function of contour level), and a Rotationally-Averaged Power Spectrum (RAPS) plot.
- If two half-maps had been deposited, the FSC curve was calculated and shown; if the depositor provided their own FSC data, these were also shown. The resulting resolution estimates at various cut-offs were summarised in a table for both curves.
- A very simple criterion was used to assess the model-to-map fit, namely the fraction of atoms that are inside the map at the depositor-recommended contour level (atom-inclusion score). This information was shown in a graph and accompanied by three orthogonal views of the superimposed map and model.
- The wwPDB validation reports contain residue-property plots for every chain in a structure. These plots highlight residues that have outliers on one or more geometric (model-only) validation criteria. In addition, residues that do not appear to fit the data very well are flagged up in a method-specific manner. For MX structures, residues with an RSRZ score greater than 2.0 are flagged (Kleywegt *et al.*, 2004). For 3DEM structures, residues with an atom-inclusion score less than 40% at the depositor-recommended contour level were flagged.

**Aims and format of the 2020 workshop.** The workshop was held at EMBL-EBI (Hinxton, UK) on 23 and 24 January 2020. The focus of the meeting was on SPA as this was and is still the most common 3DEM modality for which experimental results are deposited in EMDB and PDB. The aims of the 2020 meeting were as follows:
- To provide advice on how to improve (meta)data deposition to PDB and EMDB;
- To review the contemporaneous, preliminary 3DEM validation reports and obtain feedback and suggestions for their improvement;
- To discuss potential additional model-only, map-only and map-model validation metrics for single-particle cryoEM depositions.

The workshop was held over two days. On the first day, there were several introductory presentations outlining some of the history, explaining what information is captured during the deposition process, and describing the contemporaneous validation reports. The participants then split into three groups that each discussed the same list of topics and questions prepared by the organisers. Break-out sessions on the first day addressed data and deposition requirements, while on the second day validation metrics and reporting were discussed. The meeting concluded with reports from the three groups and a plenary discussion. The participants in the workshop are shown in **Figure 2**.

Note: in 2021 EMDB formally became a wwPDB Core Archive and EMDB as an organisation became a full partner in wwPDB ([https://www.wwpdb.org/news/news?year=2021#60d0db870882a5597783cbfe](https://www.wwpdb.org/news/news?year=2021#60d0db870882a5597783cbfe)). This means that the EMDB archive (like the PDB and BMRB archives) is now jointly managed by the five wwPDB partners (wwPDB consortium, 2024). Hence, the implementation of the recommendations of the workshop was begun collaboratively by EMDB and wwPDB staff but as of 2021 it is carried out under the aegis of wwPDB.

# Recommendations to wwPDB/EMDB on archiving and deposition

The following recommendations are mostly about improving the wwPDB deposition system OneDep, thereby improving the data content of both the PDB and EMDB archives to facilitate wider and better use of these data. Additionally, there are some recommendations that will help wwPDB engage with the 3DEM community and obtain continuous feedback for future improvements.

**Improving deposition and archive content.** The following recommendations on improving the OneDep deposition, annotation and validation system will allow for more efficient data deposition and improved data representation. The recommendations also include suggestions to improve archive content, including archiving additional metadata and requiring additional information during deposition to help with validation of maps and models.

The participants also discussed how individual PDB entries are archived and how linking between related PDB entries could be improved. This is particularly relevant as, in the case of SPA, a single imaging experiment resulting in a set of micrographs may result in identification of multiple states of macromolecules in the sample which are deposited as multiple entries, each containing a single homogeneous structure (map in EMDB or coordinate model in the PDB) with the parent publication as the only link between these related entries. In some cases, if the resulting structure models are not all described in a single publication, the link between related models deposited as different entries is lost. These related entries might contain useful biological information on, for instance, multiple compositions of a complex, multiple states of a macromolecule or complex, or ligand-bound and unbound states providing insights into active and inactive forms of the macromolecule or complex. Description of appropriate linkages between entries in different archives (PDB, EMDB, EMPIAR, SASBDB, BMRB and potentially other resources that archive experimental data used in structure determination) was also discussed as this is critical both for validation of data and models and for understanding the biology of the sample under investigation.

- **Recommendation 1.** Every 3DEM map that is represented in any manuscript figure should be deposited as a separate EMDB entry (filtered/masked as necessary).

    **Justification:** Public access to the volume data is essential for basic examination and scrutiny of what is being shown in the figure by readers and users of the structure.

- **Recommendation 2.** The deposition of unmasked, unfiltered, half-dataset reconstruction volumes (half-maps) should be made mandatory for any EMDB deposition from SPA and STA experiments. For purposes of consistent validation, EMDB should generate a raw map by averaging the two half-maps. Depositors are also encouraged to provide raw full-dataset maps (unfiltered, unmasked).

**Justification:** Half-dataset reconstruction volumes are required as input to many map and map-model validation methods.

- **Recommendation 3.** Any applied mask should be deposited and identified and annotated based on how it was used, e.g. "mask used for FSC calculation", "mask used for focused refinement", etc.

    **Justification:** Appropriate annotation of the masks is essential to ensure that they are used correctly and in the right context. The mask data is required, e.g., for reproducing FSC curves from half-maps where masks were used.

- **Recommendation 4**. Composite maps should be clearly identified as such and used in validation of the corresponding coordinate model. Each of the individual maps should be deposited and identified as a component map of the composite map.

    **Justification:** There are a growing number of entries in EMDB that are *de facto* composite maps constructed using multiple individual maps. Conventional validation techniques such as FSC cannot be applied meaningfully to composite maps and the individual maps need to be deposited to enable proper map validation.

- **Recommendation 5.** Atomic coordinates derived using 3DEM methods should be deposited to the PDB in PDBx/mmCIF format (Westbrook et al., 2005). wwPDB should provide tools to help software developers make the transition from the historic PDB format to PDBx/mmCIF.

    **Justification:** The PDBx/mmCIF format overcomes many of the shortcomings of the legacy PDB format and provides the scope and flexibility for further refinement and expansion to cover the specific needs of 3DEM.

- **Recommendation 6.** The practice of having the final map and model, related maps and masks in the same coordinate frame must continue and be enforced.

    **Justification:** Overlaying and comparing different types of structural information is a key part of analysis and validation and having this information in the same coordinate frame is therefore essential.

- **Recommendation 7.** The deposition of particle stacks should be strongly encouraged, and the OneDep system should provide a seamless mechanism for depositing these data, perhaps to EMPIAR. The associated PDB and EMDB entries should record the EMPIAR accession code assigned to the deposited data.

    **Justification:** The availability of particle stacks makes it possible to improve the validation of the corresponding 3DEM maps and atomic coordinate models. In the

future, it may also become possible to refine models directly against the 2D images.

- **Recommendation 8.** wwPDB should design a system that facilitates deposition of and access to all the relevant experimental data and structure models related to a single "investigation" or "project". The extensible PDBx/mmCIF format should be updated to define the additional semantics necessary to describe relationships between the multiple experimental data and structure models and to express the rich information.

  **Justification:** To maximise the impact of structural studies it is crucial that the rich information and biological context of relationships between different maps and models is expressed and recorded in the structure archives.

- **Recommendation 9.** wwPDB is encouraged to devise an agile mechanism to rapidly respond to developments in the 3DEM field, e.g., by forming an expert 3DEM advisory data working group. EMDB is encouraged to implement a three-tiered strategy for the dissemination of validation information which incorporates the flexibility to showcase and test the latest developments for the benefit of developers and expert users (tier 1 and 2), while only exposing those components that have gained wide acceptance and are sufficiently robust to be used by general users in the wwPDB validation reports (tier 3).

  **Justification:** As the 3DEM field continues to develop rapidly, there will be frequent and significant changes in experimental methodology and structure-determination software, as well as new developments in validation approaches. This evolution will bring with it many changes to existing data standards (e.g., controlled vocabularies to describe sample-preparation methods), new metadata requirements and new validation methodologies that may warrant incorporating into the wwPDB validation pipeline.

- **Recommendation 10.** wwPDB should continue to make validation functionality accessible via an Application Programming Interface (API). wwPDB is also encouraged to develop its validation software in such a way that it can be distributed to external users, and thus run in-house, independent of any deposition and not requiring data transfers.

  **Justification:** Providing easy access to the validation functionality and software makes it easier for external developers to integrate it into their software which not only encourages wider usage of the functionality, but has the added benefit of these external experts testing the software and providing feedback.

**Community engagement.** Engagement with a variety of stakeholder communities is an important part of wwPDB and EMDB activities and the 3DEM community will continue to benefit from such engagement. The participants felt that the following specific recommendation would help in structuring such interactions.

- **Recommendation 11.** wwPDB should organise a workshop for software developers to explore the PDBx/mmCIF extension developed for multiscale IHM models (Vallat *et al.*, 2018).

    **Justification:** The local resolution in 3DEM Coulombic potential maps often varies over the map and this impacts the precision with which a model can be built in the map. The IHM dictionary allows for (combinations of) multiscale representations including a combination of atomic coordinates and bead models (representing individual residues) or large solid volumes representing domains or complete polymer chains.

# Recommendations for validation pipeline and reports

The contents of the contemporaneous validation reports for 3DEM structures were briefly described in the introduction section. In the workshop, the various report sections were discussed by each of the three break-out groups. Many of the issues raised and suggestions made were common among the groups, suggesting that they are representative of the community's experience and opinions. In addition to reviewing the reports, potential additional model-only, map-only and model-to-map-fit validation metrics, that might be added to the validation reports in the future, were discussed. Consideration was given to global metrics (e.g., FSC-based resolution estimates), local metrics (e.g., local resolution, or local backbone normality) and metrics that can be both global and local (e.g., Ramachandran analysis identifies individual outlier residues and the overall analysis provides an outlier percentage or other score per molecule). A useful principle for wwPDB and EMDB has always been to only implement validation metrics that are well understood, widely adopted and non-controversial. For this reason, "bleeding edge" metrics are avoided until more experience has been gained with them and their applicability, performance, utility and limitations are better understood.

## -- General recommendations

Many of the recommendations pertained to improving presentation details in the contemporaneous validation reports (e.g., to show calculated and author-provided FSC curves in the same plot, and only to retain the 0.143, 0.5 and half-bit criteria). Most of these were implemented in the months following the workshop and have been available to depositors since December 2020. Several other detailed suggestions pertain to validation of models from any experimental method and will therefore require further discussion with the X-ray and NMR VTFs.

The validation reports contain an "executive summary" which includes a so-called "slider plot". This plot shows, for a few carefully selected validation criteria, how the structure compares to all structures in the PDB in terms of percentile scores. A lower score means that the structure scores worse than the bulk of the archive on that criterion (shown in red) and a higher score means that it scores better (shown in blue). Thus, these plots provide an at-a-glance overview of the quality of the structure relative to the rest of the archive (and also relative to a subset, such as all EM or all NMR structures, or all MX structures at similar resolution). It does not require knowledge of what the criteria measure or whether the values for a structure are "good" or not. They are therefore helpful both to specialists and non-specialists (e.g., referees and editors, who may not be structural biologists themselves). In the MX reports, the sliders are a mixture of model-only and model-to-data/map-fit criteria (e.g., free R-value and percentage RSRZ outliers), but the 3DEM reports include only model-based criteria. The workshop participants emphasised the importance of adding overall measures of map quality and model-to-map fit to the sliders but did not make any specific recommendations as to which criteria should be included as there is no consensus in the community yet and more research and analysis are needed.

The metrics shown in the executive summary should ideally be independent of the model parameters and restraints commonly used in refinement. The workshop participants suggested that it would be very useful to collect and report (as part of the executive summary) information about the classes of restraints used during refinement. This information would need to be reported by the refinement software and could then be harvested at deposition time. This recommendation could obviously also be implemented for MX and NMR structures and will require further discussion with the respective VTFs as well as with software developers.

**-- Model validation**

Model-validation criteria can be divided into two categories. Some criteria essentially assess how well the refinement software has been able to enforce restraints to produce a chemically and physically reasonable model. This category includes bond length and angle validation and assessment of "clashes" (unrealistically close contacts) between atoms. The other category consists of criteria that are mostly independent of the applied restraints and essentially test aspects of the model's "predictive power" (Kleywegt, 2009). This has tended to include criteria related to the main-chain and side-chain torsion-angle combinations (Ramachandran plot and rotamericity). However, certain refinement programs allow torsion-angle information to be used during refinement. Whereas this produces models with fewer outliers, these models are not necessarily better. This was first realised more than 25 years ago (Kleywegt & Jones, 1998) when such functionality had first become available in the refinement program X-PLOR (Kuszewski *et al.*, 1996) (Kuszewski *et al.*, 1997). Indeed, there are several (low-resolution) structures in the PDB that have good Ramachandran and rotamer scores, yet by other criteria are not great models. Goodhart's Law states that when a measure becomes a target, it ceases to be a good measure, the lesson here being that refinement targets should not be used as validation measures and vice versa.

To make it easier to identify cases where torsion-angle values or combinations have been restrained (or imposed, as in ideal rotamer conformations) the following recommendations were made in the workshop:

- **Recommendation 12.** Refinement software should track which types of model restraints were used and OneDep should harvest this information and display it in the executive summary of the validation report.

- **Recommendation 13.** An additional coordinate-validation metric should be included, both to validate individual residues and to present as a "slider". It is proposed to use the MolProbity CaBLAM score for this purpose (Prisant *et al.*, 2020).

Both recommendations will require additional discussions, with software providers and with other VTFs, respectively.

Additional incorporation of an appropriate model-to-map-fit criterion (see below), also as a slider, would help in identifying residues that have favourable torsion angles at the expense of their fit to the data as it is generally difficult to optimise both simultaneously (unless the map is well resolved and unambiguous).

**-- Data and map validation**

Independent of any atomic models, validation should include assessment of a number of aspects of the map and of the image data and metadata, if available. One common task, for example, is to estimate the resolution of a map, but one would also hope to detect specific pathologies, for example map anisotropy or evidence of overfitting. (In this context, "overfitting" refers to erroneous optimisation of particle orientation or other parameters due to noise in the image data, leading to deterioration of map features and often to the appearance of artefactual features in the map.) Many metrics have been proposed to quantify these and other features from the map and the data, some having gained widespread acceptance by the community, while others are still in the exploratory stages.

**Global resolution.** The community has largely settled on using the comparison of half-dataset reconstructions via the FSC plot as a useful proxy to the overall (global) resolution of a map. EMDB allowed (but did not mandate at the time) deposition of half-dataset reconstructions and FSC curves, and validation reports displayed these curves when available. It is therefore recommended that deposition of such half-dataset reconstruction volumes be made mandatory (see **Recommendation 2** above) and this was implemented in February of 2022.

**Local resolution.** If half-dataset reconstructions are available, a number of additional metrics can be deployed to characterise the maps locally (see below). However, it was felt that none of these specific algorithms had yet gained wide enough usage, or were known to be robust enough, to become part of the validation pipeline. Rather, it is recommended that candidate algorithms first be deployed and added to EMDB's Validation Analysis pages (Wang *et al.*, 2022) so that their utility and applicability can be assessed on individual structures by investigators, and analysed across the full archive by EMDB. It is expected that, over time, this will show some of these algorithms to be informative, robust and reliable enough to warrant inclusion in the wwPDB validation pipeline and reports.

Measures of local resolution such as ResMap (Kucukelbir *et al.*, 2014), BlocRes (Cardone *et al.*, 2013), MonoRes (Vilas *et al.*, 2018) and others allow a mapping of local resolution onto the 3D grid of the reconstruction, but may not be robust enough to produce comparable results in all experimental situations, or may have significant dependence on user-supplied parameter values. For example, box size or mask size parameters can significantly affect the computed resolution. A systematic, comparative study of the different local resolution metrics, enabled by routine deposition of half-dataset maps to EMDB, is recommended. In the meantime, it would be helpful to allow authors to deposit local-resolution maps as part of the deposition process.

- **Recommendation 14**. Enable deposition by users of local resolution maps.

The community consensus appears to be that estimated local-resolution values are not quantitatively reliable. Thus, the global resolution estimate could be supplemented by a coloured local-resolution map accompanied by a colour legend labelled not with specific Ångström values, but rather "better" (blue) to "worse" (red) resolution. Finally, it was recognised that for cases where an atomic model is available, a visual depiction of local resolution across the amino-acid or nucleotide sequence would be a valuable addition to future validation reports, once robust algorithms that produce such a mapping become available and are widely accepted in the community.

**Anisotropy and angular coverage.** Beyond local-resolution estimates, several measures of map anisotropy and angular coverage have been developed such as 3DFSC (Tan *et al.*, 2017), CryoEF (Naydenova & Russo, 2017), MonoDIR (Vilas *et al.*, 2020), EMDA (Warshamanage *et al.*, 2021) and SCF (Baldwin & Lyumkis, 2021). It is recommended that, once a large corpus of half-dataset maps is available in EMDB, a subset of these algorithms be systematically investigated by making them available on the EMDB Validation Analysis webpages (and future server). When community consensus has been reached, one or more of these measures could be included in the validation reports.

**Other map-only validation methods.** The validation reports already include a plot of the RAPS of the map as a function of spatial frequency. This can be useful to identify issues such as excessive filtering or sharpening. It is recommended that the validation reports include not only the RAPS of the primary map (the main deposited map that is described in the associated publication), but also (in the same plot) that of the raw map (or of the sum of the unfiltered, unsharpened, unmasked half-maps; see **Recommendation 2** above), which could help reviewers and users assess map filtering and post-processing performed by the depositors.

**Map symmetry.** The validation report should include verification that the user-supplied point-group symmetry information is correct and that the standard symmetry conventions for different point groups have been correctly followed. Symmetry information can be derived from the map by programs such as ProSHADE (Nicholls *et al.*, 2018). For large symmetric assemblies such as viruses, visual displays of the entire assembly as well as of the asymmetric unit should be included. In addition, it is important to verify that the symmetry of the deposited map matches that of the derived atomic model.

**Map-data validation metrics.** It is recommended that deposition of a stack of particle images and a minimal set of metadata to describe them be made mandatory for SPA depositions to EMDB (see **Recommendation 7** above). This will allow the development and use of many additional validation metrics, and offers substantial additional benefits, such as permitting the routine post-publication re-processing and potential improvement of structures, thereby maximising the impact of the depositors' work.

It is acknowledged that maps can be incorrectly calculated from images (van Heel, 2013) (Subramaniam, 2013) (Henderson, 2013) and that the community would benefit from validation methods that could flag such cases automatically at the time of deposition. At present, it is not feasible to deposit the raw movie files that constitute the experimental data to a central location (EMPIAR) for every EMDB deposition. Moreover, a raw data set may give rise to multiple EMDB entries due to different compositional or conformational states. However, the subset of boxed particles used to create a map constitutes the raw data for a given entry and is sufficient to reproduce the map. Thus, wide availability of particle stacks would allow implementation of map-validation tools, and they should be deposited in conjunction with metadata containing single-particle parameters describing the exact relationship between the map and image data, which will enable assessment of the reliability of parameters such as angular uncertainty or estimation of map overfitting. It is recommended, following best practises regarding data and metadata formatting and conventions (likely following recommendations of (Marabini *et al.*, 2016), and choosing widely-used formats such as MRC/CCP4 and Star or XML files), that deposition of image-stack data and metadata becomes possible (and later mandatory).

It was recognised that data and map validation will remain a field of active research for some time, with new algorithms being proposed and released regularly. Hence, for many of these tasks and validation methods the community has not yet converged on specific solutions that could be said to have become widely accepted standards. For this reason, further research, including comparative and archive-wide studies, is encouraged, and could be facilitated by EMDB or EMPIAR.

-- **Model-to-map fit validation**

Going beyond assessment of the quality of data/map and model separately, a crucial part of validation is assessing how well the model fits the data/map. In MX this is typically accomplished with global reciprocal-space measures such as the R-value and Rfree-value (Brünger, 1992), and locally through real-space statistics such as the per-residue Real-Space R-value (RSR) (Jones *et al.*, 1991), Real-Space Correlation Coefficient (RSCC) (Jones *et al.*, 1991), and RSRZ scores (Kleywegt *et al.*, 2004). Individual residues, ligands, etc. that are outliers on such real-space measures ought to be inspected to ascertain whether they are due to a locally poor map, or constitute a poorly built part of the model, or possibly both. Local model rebuilding and further refinement may be able to improve the fit prior to publication and deposition. Any outliers remaining in the final model should be flagged up to depositors and users of the archives.

In the case of SPA 3DEM, a map-model FSC plot provides an indication of the correlation between the experimental map and one computed from the model as a function of resolution.

- **Recommendation 15.** It is recommended that such a plot be calculated (ideally in a way similar to the half-map FSC calculation) and included in the validation reports, along with the resolution value at which FSC=0.5. As for half-map FSC calculations, several parameters influence the outcome of the computation (e.g., map masking, model-to-map simulation parameters) and there is no clearly preferred, fully-unsupervised software solution as yet. It is recommended that existing popular solutions (e.g., Phenix mtriage) be implemented in the EMDB Validation Analysis webpages. In instances where a customised mask volume has been used for resolution estimation, the noise-substitution-corrected FSC curve (Chen *et al.*, 2013) should also be plotted.

In recent years, quite a few metrics comparable to the real-space measures for MX have been developed for 3DEM, including EMringer (Barad *et al.*, 2015), SMOC (Joseph *et al.*, 2016), Q-score (Pintilie *et al.*, 2020) and CCC (Warshamanage *et al.*, 2021). In the current validation reports, the per-residue atom-inclusion score is used, which is defined as the fraction of atoms of a residue that lie within the map if it is contoured at the depositor-recommended level. There are several issues with this score: it depends on the composition of the sample (e.g., proteins, nucleic acids, lipid membranes) and the subjective choice of a single contour level and there also appears to be a resolution-dependence (Lawson *et al.*, 2021). Moreover, it may be tempting to "optimise" (i.e., unduly lower) the recommended contour level so as to maximise the atom-inclusion score, which is obviously counter-productive. It was generally agreed that more experience with these metrics is required before any of them can be recommended to replace the atom-inclusion score. EMDB will incorporate a number of these scores into their Validation Analysis pages. This will enable individuals to inspect the metrics' behaviour for structures that they are familiar with and will also allow archive-wide analysis and comparison of these metrics.

A need was identified for methods to calculate 3D difference maps between map and model and for ways of analysing these automatically. In addition, more understanding and experience is needed of the relationship of model temperature factors and map characteristics such as local resolution before any recommendations can be made about their validation. (In MX validation reports, anisotropic atomic-displacement parameters are currently not validated.)

The molecular weight calculated from a plot of enclosed volume as a function of contour level should be compared to the reported molecular weight, both as absolute values and to indicate relative proportion. A related parameter that could be reported is the ratio of the surface area to the enclosed map volume to provide a measure of the level of detail of the map.

Finally, an interesting suggestion was made to provide a visual illustration of representative model-to-map fit in both a relatively good and a relatively poor part of the map where a model has been constructed (*e.g.*, three orthogonal views of map and model). The good and poor regions should be small (a few residues) and they could either be designated by the authors or identified automatically (*e.g.*, the three consecutive residues with the highest and lowest average model-to-map-fit score, respectively).

## Recommendations and challenges for software and methods developers

The discussions in the workshop led to several recommendations for developers of both data-processing and model-refinement software. One overarching theme that emerged was the need for developers to work with wwPDB to collate metadata in files so that they can be harvested automatically at deposition time.

The workshop further identified several unsolved questions and challenges where further research, methods development and analysis are needed:

1. For both map and data validation, are there any candidate criteria that might be suitable for inclusion in the "slider" graphs in the validation reports? Such criteria should be well-tested and their behaviour (*e.g.*, in relation to quality and dependence on resolution) well understood. They should also not be closely related to parameters that are directly refined or optimised in the structure-determination process, and not be easy to "fudge".

2. An open question at this time is how 2D raw data (*e.g.*, particle images) can best be used to validate the 3D map.

3. In the area of model-to-map-fit assessment, a "slider criterion" is urgently needed as well. Furthermore, robust tools to compute difference maps between data and models need to be developed.

4. Although older and lower-resolution cryoEM models may contain (parts with) only Cα atoms, there are at present few if any validation methods for such and other coarse-grained models (Kleywegt, 1997), so it would be helpful if these were developed.

5. A method to provide an unbiased optimal contour level for map viewing and inclusion-score calculation (global and at local levels, *e.g.*, per domain) would be extremely useful.

6. Methods to assess if structural features observed at a given resolution are commensurate with expectations or experience need to be developed. Machine-learning approaches might be suitable to address this problem.

Methods developers in the cryoEM and related fields (*e.g.*, X-ray crystallography) are encouraged to address these challenges. As methods gain acceptance in the field, a selected subset could be added to the Validation Analysis web pages of EMDB, ahead of eventual addition of those that are proven to be robust and informative to the wwPDB validation reports.

## Considerations for the community

We recognise that it is important not to be overly prescriptive about the experimental and computational practises employed in map and model generation. The aim of the workshop and this white paper is to highlight tools that can flag outliers (which in turn may be used to identify errors) and minimise over-interpretation of results, and to support wwPDB in their goal of improving support of the development and widespread use of validation methods. It is also important to emphasise that the aim of the validation exercise is not to help authors obtain a model that has no outliers on any given measures (such as Ramachandran analysis). Instead, the goal is to help structural biologists identify potential issues in a reported model or in the underlying data, so that they may address these through remodelling or reprocessing, and deposit for public use a final model that is a more faithful interpretation of the experimental data and that also incorporates appropriate prior knowledge (*e.g.*, chemical geometry or noise statistics) as accurately as possible. Finally, validation reports should help users of these structures to identify features of a structural model which are possibly more, or less, reliable than others, and to compare multiple available models to identify those that are best suited for their specific applications.

# Way forward

Shortly after the workshop, EMDB and wwPDB staff began implementing many of the recommendations concerning the 3DEM validation pipeline and reports. Their efforts have resulted in an updated software pipeline with which present-day validation reports are generated for 3DEM structures in PDB and EMDB. To assist in the process of assessing the applicability, performance and limitations of multiple alternative validation methods (*e.g.*, to assess model-to-map fit), it was agreed that such methods should first be implemented as part of the EMDB Validation Analysis web pages (and future server), as recommended by the workshop. In this way, experts can check how these methods perform and compare them using cases that they are familiar with. Moreover, it will enable archive-wide analyses and together these will inform future recommendations. An overview of the implementation of the multi-tiered approach to validation recommended by the workshop is provided in (Wang *et al.*, 2022). Several of the recommended validation features and metrics have since been made available through the EMDB Validation Analysis resource ([https://emdb-empiar.org/va](https://emdb-empiar.org/va)) and the wwPDB validation pipeline and reports.

The 2020 workshop focussed on SPA structure determination. However, many of the recommendations also apply to other 3DEM modalities (*e.g.*, STA). In the future, additional specialist workshops may be held for other modalities (*e.g.*, cellular tomography). It should be noted that both the EMDB Validation Analysis resource (Wang *et al.*, 2022) and the wwPDB validation pipeline can assess 3DEM volumes for all EM modalities supported by EMDB, and regardless of whether or not there is a model, albeit that the amount of validation information provided may be limited.

It should also be noted that the workshop was held a year and a half before reliable predicted protein structures became available at proteome scale (Tunyasuvunakool *et al.*, 2021). These predicted structures are showing great promise to assist in experimental structure determination (Terwilliger *et al.*, 2023). However, large-scale deposition of experimental models based on predicted structures may require development of novel methods for structure validation as the prediction software has been trained on the contents of the PDB and may thus reproduce proper packing etc. (Jumper *et al.*, 2021).

The recommendations made in this paper reflect the insights and needs of the community at the time of the workshop. Many of them have already been implemented in the validation reports and in EMDB policies and resources. The recommendations have continued and will continue to evolve with the science and they continue to be refined in close consultation with the community. In 2024, wwPDB will convene a working group to advise it on cryoEM data deposition and validation. This working group will communicate by email and through regular online meetings and can thus provide feedback and advice at relatively short notice. Improving the 3DEM validation reports and addressing unresolved issues (*e.g.*, which criteria to use as "sliders" and which metrics to use to assess model-to-map fit) are currently being addressed. The reports will evolve over time as the methodology advances, more experience and insight are gained, and consensus recommendations materialise. The wider 3DEM community is encouraged to discuss issues of data deposition and validation in a variety of contexts, *e.g.*, in national and international meetings, on bulletin boards and mailing lists, around the water-cooler, on social media, *etc*. and unsolicited advice is welcomed by wwPDB through its help desk or in person.

## Acknowledgements

We would like to thank Professor Kaoru Mitsuoka (Osaka University) for his contributions to the discussions in the workshop. The workshop was supported by funding to PDBe and EMDB by the Wellcome Trust (grant 104948/Z/14/Z to GJK, SV and AP) and by the European Molecular Biology Laboratory. Travel was supported by PDBe, EMDB, RCSB PDB, PDBj, BMRB and EMDR. RCSB PDB is jointly funded by the National Science Foundation (DBI-1832184), the US Department of Energy (DE-SC0019749), and the National Cancer Institute, National Institute of Allergy and Infectious Diseases, and National Institute of General Medical Sciences of the National Institutes of Health under grant R01GM133198. PDBj is funded by JST-NBDC and BMRB by the National Institute of General Medical Sciences of the National Institutes of Health under grant R24GM150793. EMDR was funded by the National Institute of General Medical Sciences of the National Institutes of Health under grant R01GM079429. The organisers wish to thank Irina Beszonova (UConn Health, USA) for providing the graphical abstract image, and Pauline Haslam and Roisin Dunlop (EMBL-EBI) for logistical support.

The Article-Processing Costs were funded by EMBL.

# References

Abdul Ajees, A., Gunasekaran, K., Volanakis, J. E., Narayana, S. V. L., Kotwal, G. J. & Murthy, H. M. K. (2006). *Nature* **444**, 221–225.

Baldwin, P. R. & Lyumkis, D. (2021). *Prog. Biophys. Mol. Biol.* **160**, 53–65.

Barad, B. A., Echols, N., Wang, R. Y.-R., Cheng, Y., DiMaio, F., Adams, P. D. & Fraser, J. S. (2015). *Nat. Methods* **12**, 943–946.

Berman, H., Henrick, K. & Nakamura, H. (2003). *Nat. Struct. Biol.* **10**, 980.

Borrell, B. (2009). *Nature* **462**, 970.

Brändén, C.-I. & Jones, T. A. (1990). *Nature* **343**, 687–689.

Brünger, A. T. (1992). *Nature* **355**, 472–475.

Cardone, G., Heymann, J. B. & Steven, A. C. (2013). *J. Struct. Biol.* **184**, 226–236.

Chen, S., McMullan, G., Faruqi, A. R., Murshudov, G. N., Short, J. M., Scheres, S. H. W. & Henderson, R. (2013). *Ultramicroscopy* **135**, 24–35.

Davis, I. W., Murray, L. W., Richardson, J. S. & Richardson, D. C. (2004). *Nucleic Acids Res.* **32**, W615–W619.

Gore, S., Sanz García, E., Hendrickx, P. M. S., Gutmanas, A., Westbrook, J. D., Yang, H., Feng, Z., Baskaran, K., Berrisford, J. M., Hudson, B. P., Ikegawa, Y., Kobayashi, N., Lawson, C. L., Mading, S., Mak, L., Mukhopadhyay, A., Oldfield, T. J., Patwardhan, A., Peisach, E., Sahni, G., Sekharan, M. R., Sen, S., Shao, C., Smart, O. S., Ulrich, E. L., Yamashita, R., Quesada, M., Young, J. Y., Nakamura, H., Markley, J. L., Berman, H. M., Burley, S. K., Velankar, S. & Kleywegt, G. J. (2017). *Structure* **25**, 1916–1927.

Gore, S., Velankar, S. & Kleywegt, G. J. (2012). *Acta Crystallogr. D Biol. Crystallogr.* **68**, 478–483.

van Heel, M. (2013). *Proc. Natl. Acad. Sci. U. S. A.* **110**, E4175–E4177.

Henderson, R. (2013). *Proc. Natl. Acad. Sci. U. S. A.* **110**, 18037–18041.

Henderson, R., Baldwin, J. M., Ceska, T. A., Zemlin, F., Beckmann, E. & Downing, K. H. (1990). *J. Mol. Biol.* **213**, 899–929.

Henderson, R., Sali, A., Baker, M. L., Carragher, B., Devkota, B., Downing, K. H., Egelman, E. H., Feng, Z., Frank, J., Grigorieff, N., Jiang, W., Ludtke, S. J., Medalia, O., Penczek, P. A., Rosenthal, P. B., Rossmann, M. G., Schmid, M. F., Schröder, G. F., Steven, A. C., Stokes, D. L., Westbrook, J. D., Wriggers, W., Yang, H., Young, J., Berman, H. M., Chiu, W., Kleywegt, G. J. & Lawson, C. L. (2012). *Structure* **20**, 205–214.

Hooft, R. W., Vriend, G., Sander, C. & Abola, E. E. (1996). *Nature* **381**, 272.

Iudin, A., Korir, P. K., Salavert-Torres, J., Kleywegt, G. J. & Patwardhan, A. (2016). *Nat. Methods* **13**, 387–388.

Iudin, A., Korir, P. K., Somasundharam, S., Weyand, S., Cattavitello, C., Fonseca, N., Salih, O., Kleywegt, G. J. & Patwardhan, A. (2023). *Nucleic Acids Res.* **51**, D1503–D1511.

Jones, T. A., Zou, J. Y., Cowan, S. W. & Kjeldgaard, M. (1991). *Acta Crystallogr. A* **47 ( Pt 2)**, 110–119.

Joseph, A. P., Malhotra, S., Burnley, T., Wood, C., Clare, D. K., Winn, M. & Topf, M. (2016). *Methods* **100**, 42–49.

Jumper, J., Evans, R., Pritzel, A., Green, T., Figurnov, M., Ronneberger, O., Tunyasuvunakool, K., Bates, R., Žídek, A., Potapenko, A., Bridgland, A., Meyer, C., Kohl, S. A. A., Ballard, A. J., Cowie, A., Romera-Paredes, B., Nikolov, S., Jain, R., Adler, J., Back, T., Petersen, S., Reiman, D., Clancy, E., Zielinski, M., Steinegger, M., Pacholska, M., Berghammer, T., Bodenstein, S., Silver, D., Vinyals, O., Senior, A. W., Kavukcuoglu, K., Kohli, P. & Hassabis, D. (2021). *Nature* **596**, 583–589.

Kleywegt, G. J. (1997). *J. Mol. Biol.* **273**, 371–376.

Kleywegt, G. J. (2009). *Acta Crystallogr. D Biol. Crystallogr.* **65**, 134–139.

Kleywegt, G. J., Harris, M. R., Zou, J. Y., Taylor, T. C., Wählby, A. & Jones, T. A. (2004). *Acta Crystallogr. D Biol. Crystallogr.* **60**, 2240–2249.

Kleywegt, G. J. & Jones, T. A. (1995). *Structure* **3**, 535–540.

Kleywegt, G. J. & Jones, T. A. (1997). *Methods Enzymol.* **277**, 208–230.

Kleywegt, G. J. & Jones, T. A. (1998). *Acta Crystallogr. D Biol. Crystallogr.* **54**, 1119–1131.

Kucukelbir, A., Sigworth, F. J. & Tagare, H. D. (2014). *Nat. Methods* **11**, 63–65.

Kühlbrandt, W. (2014). *Science* **343**, 1443–1444.

Kuszewski, J., Gronenborn, A. M. & Clore, G. M. (1996). *Protein Sci.* **5**, 1067–1080.

Kuszewski, J., Gronenborn, A. M. & Clore, G. M. (1997). *J. Magn. Reson.* **125**, 171–177.

Lagerstedt, I., Moore, W. J., Patwardhan, A., Sanz-García, E., Best, C., Swedlow, J. R. & Kleywegt, G. J. (2013). *J. Struct. Biol.* **184**, 173–181.

Laskowski, R. A., MacArthur, M. W., Moss, D. S. & Thornton, J. M. (1993). *J. Appl. Crystallogr.* **26**, 283–291.

Lawson, C. L., Berman, H. M. & Chiu, W. (2020). *Struct Dyn* **7**, 014701.

Lawson, C. L. & Chiu, W. (2018). *J. Struct. Biol.* **204**, 523–526.

Lawson, C. L., Kryshtafovych, A., Adams, P. D., Afonine, P. V., Baker, M. L., Barad, B. A., Bond, P., Burnley, T., Cao, R., Cheng, J., Chojnowski, G., Cowtan, K., Dill, K. A., DiMaio, F., Farrell, D. P., Fraser, J. S., Herzik, M. A., Jr, Hoh, S. W., Hou, J., Hung, L.-W., Igaev, M., Joseph, A. P., Kihara, D., Kumar, D., Mittal, S., Monastyrskyy, B., Olek, M., Palmer, C. M., Patwardhan, A., Perez, A., Pfab, J., Pintilie, G. D., Richardson, J. S., Rosenthal, P. B., Sarkar, D., Schäfer, L. U., Schmid, M. F., Schröder, G. F., Shekhar, M., Si, D., Singharoy, A., Terashi, G., Terwilliger, T. C., Vaiana, A., Wang, L., Wang, Z., Wankowicz, S. A., Williams, C. J., Winn, M., Wu, T., Yu, X., Zhang, K., Berman, H. M. & Chiu, W. (2021). *Nat. Methods* **18**, 156–164.

Ludtke, S. J., Lawson, C. L., Kleywegt, G. J., Berman, H. & Chiu, W. (2012). *Biopolymers* **97**, 651–654.

Marabini, R., Carragher, B., Chen, S., Chen, J., Cheng, A., Downing, K. H., Frank, J., Grassucci, R. A., Bernard Heymann, J., Jiang, W., Jonic, S., Liao, H. Y., Ludtke, S. J., Patwari, S., Piotrowski, A. L., Quintana, A., Sorzano, C. O. S., Stahlberg, H., Vargas, J., Voss, N. R., Chiu, W. & Carazo, J. M. (2015). *J. Struct. Biol.* **190**, 348–359.

Marabini, R., Ludtke, S. J., Murray, S. C., Chiu, W., de la Rosa-Trevín, J. M., Patwardhan, A., Heymann, J. B. & Carazo, J. M. (2016). *J. Struct. Biol.* **194**, 156–163.

Montelione, G. T., Nilges, M., Bax, A., Güntert, P., Herrmann, T., Richardson, J. S., Schwieters, C. D., Vranken, W. F., Vuister, G. W., Wishart, D. S., Berman, H. M., Kleywegt, G. J. & Markley, J. L. (2013). *Structure* **21**, 1563–1570.

Naydenova, K. & Russo, C. J. (2017). *Nat. Commun.* **8**, 629.

Nicholls, R. A., Tykac, M., Kovalevskiy, O. & Murshudov, G. N. (2018). *Acta Crystallogr D Struct Biol* **74**, 492–505.

Patwardhan, A., Ashton, A., Brandt, R., Butcher, S., Carzaniga, R., Chiu, W., Collinson, L., Doux, P., Duke, E., Ellisman, M. H., Franken, E., Grünewald, K., Heriche, J.-K., Koster, A., Kühlbrandt, W., Lagerstedt, I., Larabell, C., Lawson, C. L., Saibil, H. R., Sanz-García, E., Subramaniam, S., Verkade, P., Swedlow, J. R. & Kleywegt, G. J. (2014). *Nat. Struct. Mol. Biol.* **21**, 841–845.

Patwardhan, A., Carazo, J.-M., Carragher, B., Henderson, R., Heymann, J. B., Hill, E., Jensen, G. J., Lagerstedt, I., Lawson, C. L., Ludtke, S. J., Mastronarde, D., Moore, W. J., Roseman, A., Rosenthal, P., Sorzano, C.-O. S., Sanz-García, E., Scheres, S. H. W., Subramaniam, S., Westbrook, J., Winn, M., Swedlow, J. R. & Kleywegt, G. J. (2012). *Nat. Struct. Mol. Biol.* **19**, 1203–1207.

Pintilie, G., Zhang, K., Su, Z., Li, S., Schmid, M. F. & Chiu, W. (2020). *Nat. Methods* **17**, 328–334.

Prisant, M. G., Williams, C. J., Chen, V. B., Richardson, J. S. & Richardson, D. C. (2020). *Protein Sci.* **29**, 315–329.

Read, R. J., Adams, P. D., Arendall, W. B., 3rd, Brunger, A. T., Emsley, P., Joosten, R. P., Kleywegt, G. J., Krissinel, E. B., Lütteke, T., Otwinowski, Z., Perrakis, A., Richardson, J. S., Sheffler, W. H., Smith, J. L., Tickle, I. J., Vriend, G. & Zwart, P. H. (2011). *Structure* **19**, 1395–1412.

Subramaniam, S. (2013). *Proc. Natl. Acad. Sci. U. S. A.* **110**, E4172–E4174.

Tagari, M., Newman, R., Chagoyen, M., Carazo, J. M. & Henrick, K. (2002). *Trends Biochem. Sci.* **27**, 589.

Tan, Y. Z., Baldwin, P. R., Davis, J. H., Williamson, J. R., Potter, C. S., Carragher, B. & Lyumkis, D. (2017). *Nat. Methods* **14**, 793–796.

Terwilliger T. C., Liebschner D., Croll T. I., Williams C. J., McCoy A. J., Poon B. K., Afonine P. V., Oeffner R. D., Richardson J. S., Read R. J. & Adams P. D. (2023). *Nat. Methods*, in press (DOI 10.1038/s41592-023-02087-4).

Tunyasuvunakool, K., Adler, J., Wu, Z., Green, T., Zielinski, M., Žídek, A., Bridgland, A., Cowie, A., Meyer, C., Laydon, A., Velankar, S., Kleywegt, G. J., Bateman, A., Evans, R., Pritzel, A., Figurnov, M., Ronneberger, O., Bates, R., Kohl, S. A. A., Potapenko, A., Ballard, A. J., Romera-Paredes, B., Nikolov, S., Jain, R., Clancy, E., Reiman, D., Petersen, S., Senior, A. W., Kavukcuoglu, K., Birney, E., Kohli, P., Jumper, J. & Hassabis, D. (2021). *Nature* **596**, 590–596.

Vallat, B., Webb, B., Westbrook, J. D., Sali, A. & Berman, H. M. (2018). *Structure* **26**, 894–904.e2.

Vilas, J. L., Gómez-Blanco, J., Conesa, P., Melero, R., Miguel de la Rosa-Trevín, J., Otón, J., Cuenca, J., Marabini, R., Carazo, J. M., Vargas, J. & Sorzano, C. O. S. (2018). *Structure* **26**, 337–344.e4.

Vilas, J. L., Tagare, H. D., Vargas, J., Carazo, J. M. & Sorzano, C. O. S. (2020). *Nat. Commun.* **11**, 55.

Wang, Z., Patwardhan, A. & Kleywegt, G. J. (2022). *Acta Crystallogr D Struct Biol* **78**, 542–552.

Warshamanage, R., Yamashita, K. & Murshudov, G. N. (2021). *J. Struct. Biol.* **214**, 107826.

Westbrook, J., Henrick, K., Ulrich, E. L., Berman, H. M. (2005). "The Protein Data Bank exchange dictionary" in International Tables for Crystallography G. Definition and exchange of crystallographic data, S.R. Hall and B. McMahon, Editors, Springer: Dordrecht, The Netherlands. p. 195-198.

wwPDB consortium (2019). *Nucleic Acids Res.* **47**, D520–D528.

wwPDB consortium (2024). *Nucleic Acids Res.,* **52**, D456-D465.

Young, J. Y., Westbrook, J. D., Feng, Z., Sala, R., Peisach, E., Oldfield, T. J., Sen, S., Gutmanas, A., Armstrong, D. R., Berrisford, J. M., Chen, L., Chen, M., Di Costanzo, L., Dimitropoulos, D., Gao, G., Ghosh, S., Gore, S., Guranovic, V., Hendrickx, P. M. S., Hudson, B. P., Igarashi, R., Ikegawa, Y., Kobayashi, N., Lawson, C. L., Liang, Y., Mading, S., Mak, L., Mir, M. S., Mukhopadhyay, A., Patwardhan, A., Persikova, I., Rinaldi, L., Sanz-Garcia, E., Sekharan, M. R., Shao, C., Swaminathan, G. J., Tan, L., Ulrich, E. L., van Ginkel, G., Yamashita, R., Yang, H., Zhuravleva, M. A., Quesada, M., Kleywegt, G. J., Berman, H. M., Markley, J. L., Nakamura, H., Velankar, S. & Burley, S. K. (2017). *Structure* **25**, 536–545.

Zhu, Y., Carragher, B., Glaeser, R. M., Fellmann, D., Bajaj, C., Bern, M., Mouche, F., de Haas, F., Hall, R. J., Kriegman, D. J., Ludtke, S. J., Mallick, S. P., Penczek, P. A., Roseman, A. M., Sigworth, F. J., Volkmann, N. & Potter, C. S. (2004). *J. Struct. Biol.* **145**, 3–14.

**Corresponding author:**

G.J. Kleywegt, European Molecular Biology Laboratory, European Bioinformatics Institute (EMBL-EBI), Wellcome Genome Campus, Hinxton, Cambridgeshire, CB10 1SD, United Kingdom. Email: gerard@ebi.ac.uk

**Authors:**

**(Note: all author details have been provided in a separate Excel file.)**

Gerard J. Kleywegt [1], Paul D. Adams [2], Sarah J. Butcher [3], Catherine L. Lawson [4], Alexis Rohou [5], Peter B. Rosenthal [6], Sriram Subramaniam [7], Maya Topf [8], Sanja Abbott [1], Philip R. Baldwin [9], John M. Berrisford [1], Gérard Bricogne [10], Preeti Choudhary [1], Tristan I. Croll [11], Radostin Danev [12], Sai J. Ganesan [13], Timothy Grant [14], Aleksandras Gutmanas [1], Richard Henderson [15], J. Bernard Heymann [16], Juha T. Huiskonen [3], Andrei Istrate [1], Takayuki Kato [17], Gabriel C. Lander [18], Shee-Mei Lok [19], Steven J. Ludtke [20], Garib N. Murshudov [15], Ryan Pye [1], Grigore D. Pintilie [21], Jane S. Richardson [22], Carsten Sachse [23], Osman Salih [1], Sjors H.W. Scheres [15], Gunnar F. Schroeder [23], Carlos Oscar S. Sorzano [24], Scott M. Stagg [25], Zhe Wang [1], Rangana Warshamanage [15], John D. Westbrook (+) [4], Martyn D. Winn [26], Jasmine Y. Young [4], Stephen K. Burley [4], Jeffrey C. Hoch [27], Genji Kurisu [17], Kyle Morris [1], Ardan Patwardhan [1], Sameer Velankar [1].

(+) Deceased.

**Affiliations:**

[1] EMBL-EBI, Cambridge, UK
[2] Lawrence Berkeley Laboratory, Berkeley, CA, USA and University of California, Berkeley, CA, USA
[3] University of Helsinki, Helsinki, Finland
[4] RCSB Protein Data Bank, Rutgers, The State University of New Jersey, USA
[5] Genentech, South San Francisco, California, USA
[6] The Francis Crick Institute, London, UK
[7] University of British Columbia, Vancouver, BC, Canada
[8] Birkbeck, University of London, London, UK
[9] The Salk Institute for Biological Studies, La Jolla, CA, USA
[10] Global Phasing Limited, Cambridge, UK
[11] University of Cambridge, Cambridge, UK
[12] The University of Tokyo, Tokyo, Japan
[13] University of California at San Francisco, San Francisco, CA, USA
[14] Morgridge Institute for Research, Madison, WI, USA

[15] MRC Laboratory of Molecular Biology, Cambridge, UK
[16] National Institutes of Health, Bethesda, MD, USA
[17] Osaka University, Suita, Osaka, Japan
[18] Scripps Research, La Jolla, CA, USA
[19] Duke-NUS Medical School, Singapore
[20] Baylor College of Medicine, Houston, TX, USA
[21] Stanford University, Stanford, CA, USA
[22] Duke University, Durham, NC, USA
[23] Forschungszentrum Jülich, Germany
[24] National Center of Biotechnology (CSIC), Madrid, Spain
[25] Florida State University, Tallahassee, FL, USA
[26] Science and Technology Facilities Council, Research Complex at Harwell, Oxon, UK
[27] UConn Health, USA

**Present addresses:**

**Baldwin**: Baylor College of Medicine, Houston, TX, USA
**Berrisford**: AstraZeneca, Cambridge, UK
**Croll**: Altos Labs, Cambridge, UK
**Gutmanas**: Wren Therapeutics Ltd, Cambridge, UK
**Heymann**: Frederick National Laboratory, Frederick, MD, USA
**Istrate**: EDS Bioinformatics, Lonza Biologics, Little Chesterford, UK
**Topf**: Leibniz Institute for Experimental Virology (HPI) and Universitätsklinikum Hamburg Eppendorf (UKE), Centre for Structural Systems Biology (CSSB), Hamburg, Germany
**Warshamanage**: Science and Technology Facilities Council, Research Complex at Harwell, Oxon, UK

**Figure captions**

**Figure 1**. The "*resolution revolution*" was largely catalysed by new detector technologies in EM that had particular utility in cryoEM (Kühlbrandt, 2014). These developments have in subsequent years been bolstered by software developers, instrument manufacturers and the complementary investment of a large part of the scientific workforce utilising cryoEM. All together this has generated a sustained and approximately logarithmic growth in the number of 3DEM depositions to EMDB (panel a). Further, the technological improvements to the SPA workflow in particular have resulted in a steady increase in the number of 3DEM maps determined and deposited at resolutions sufficient for building an atomic model, in particular <3 and 3-4 Å (panel b). This is further reflected in the number of structures based on cryoEM data deposited to the PDB (panel c).

(a) The number of released EMDB entries (on a logarithmic scale), per year (blue) and cumulatively (orange). Data as of December 2023 from: https://www.ebi.ac.uk/emdb/statistics/emdb_entries_year.
(b) The number of released EMDB entries per year in a number of resolution bins, from 2010 until December 2023 (data from: https://www.ebi.ac.uk/emdb/statistics/emdb_resolution_trends_2).
(c) The annually released (dark blue) and cumulative (light blue) number of EM-based structures in the PDB as a function of year, from 2010 until December 2023 (data from: https://www.rcsb.org/stats/growth/growth-em).

**Figure 2**. The participants in the workshop (not in the photo: S. Abbott and S.J. Ganesan).

**Figure 1 (panel a).**

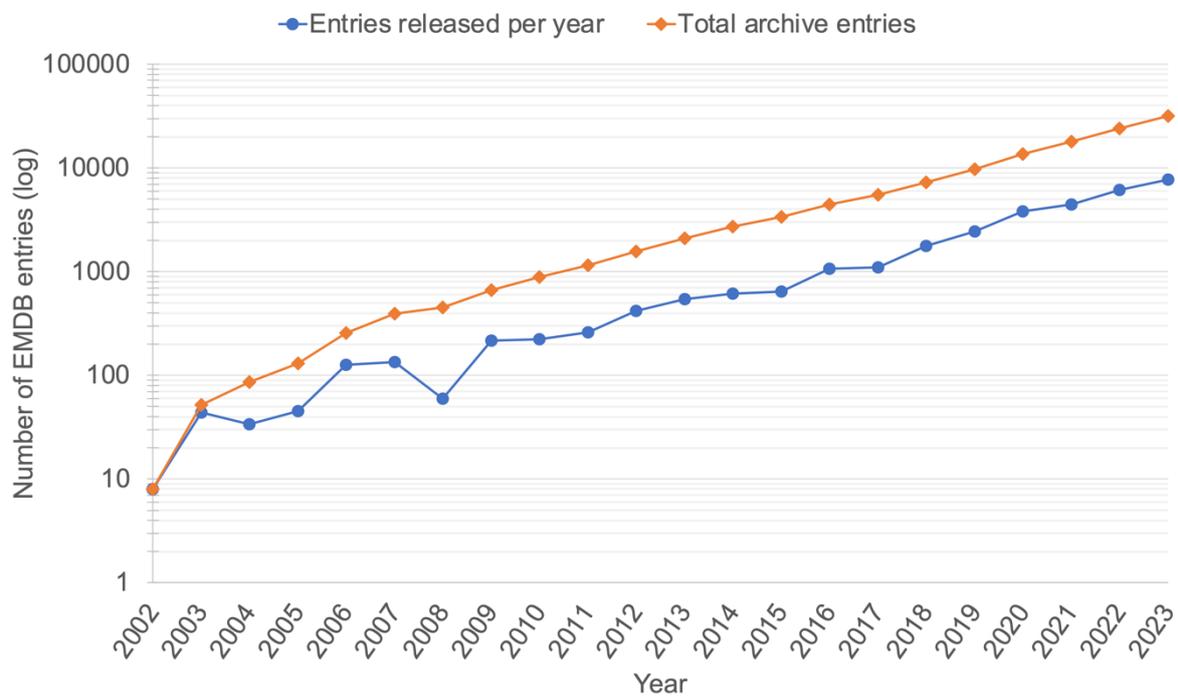

**Figure 1 (panel b).**

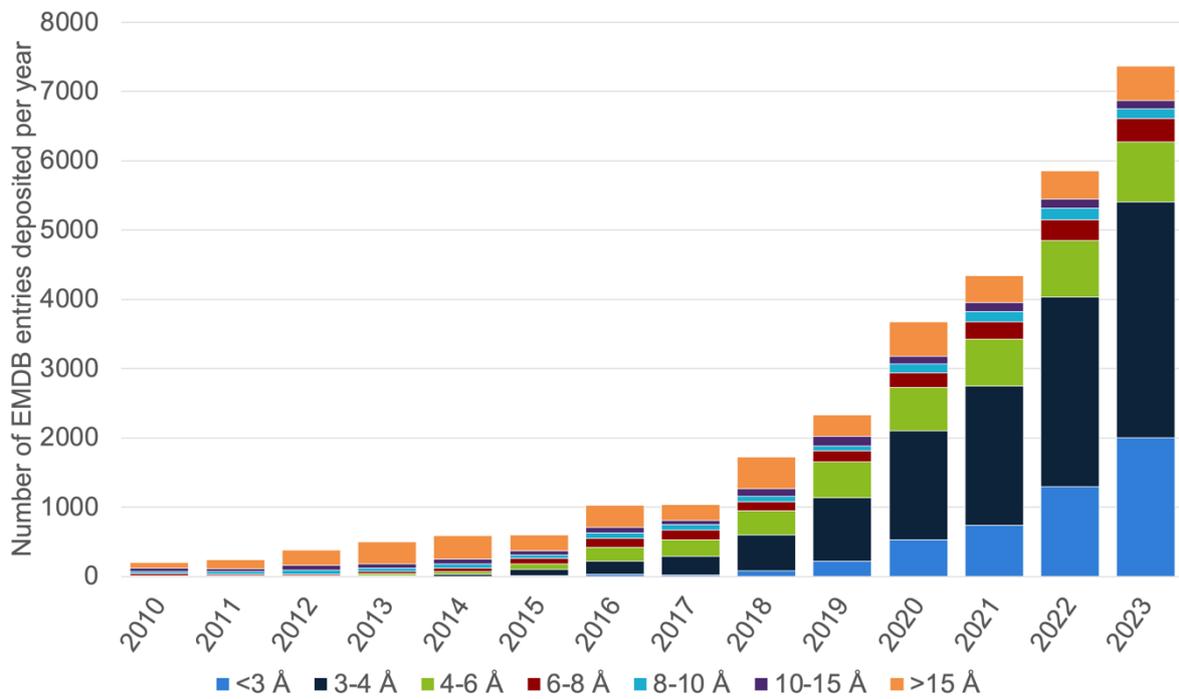

**Figure 1 (panel c).**

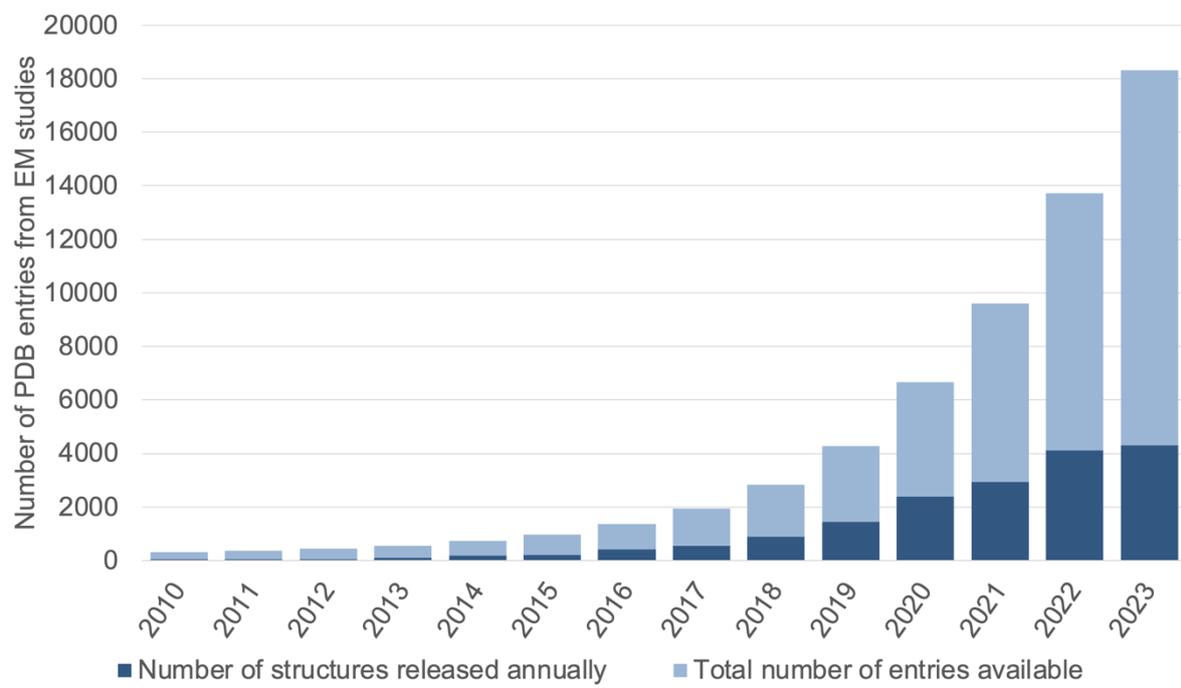

**Figure 2**.

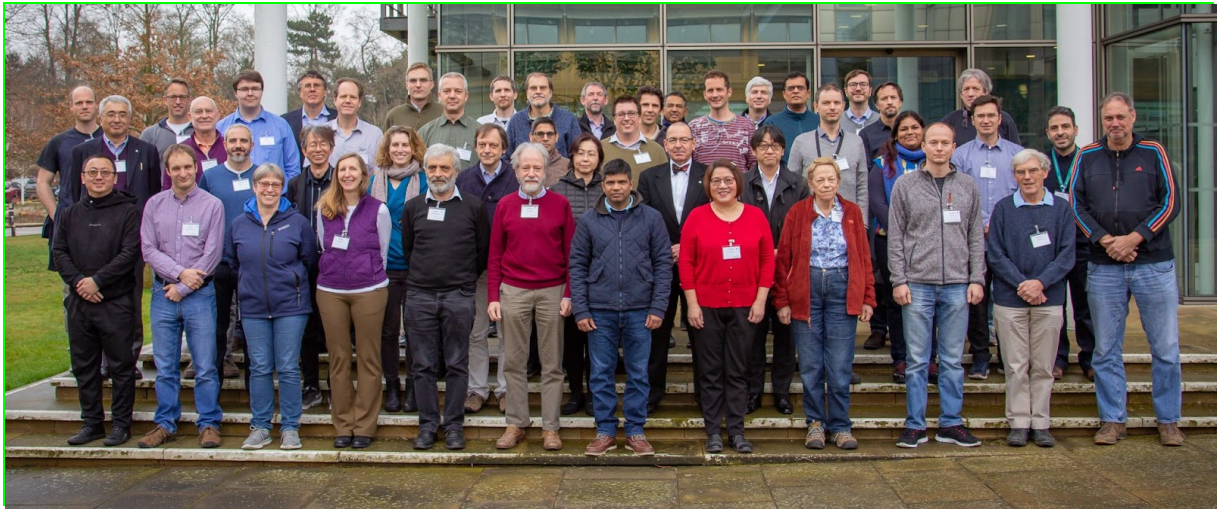